\documentclass[num-refs]{wiley-article}

\usepackage{siunitx}
\usepackage{amsmath, amssymb}
\usepackage{booktabs}
\usepackage{multirow}
\usepackage{tabularx}
\usepackage{array}
\usepackage{adjustbox}
\usepackage{url}
\usepackage{graphicx}
\usepackage{environ}
\usepackage{subcaption}
\usepackage{microtype}
\usepackage{enumitem}
\usepackage{placeins}
\usepackage{balance}
\usepackage[ruled,vlined,linesnumbered]{algorithm2e} 
\usepackage{xcolor}

\papertype{Original Article}

\title{AgentR: A Stateful and Recovery-Aware Software Architecture for LLM-based Auditable Workflows}

\abbrevs{ACID, atomicity, consistency, isolation, durability; API, application programming interface; LLM, large language model; RAG, retrieval-augmented generation; REST, representational state transfer.}

\author[1\authfn{1}]{Riya Samanta}
\author[2\authfn{1}]{Bidyut Saha}
\author[3]{Soumya Kanti Ghosh}
\author[4]{Rajkumar Buyya}

\contrib[\authfn{1}]{Equally contributing authors.}

\affil[1]{Department of Computer Science \& Engineering, Techno India University, West Bengal, India}
\affil[2] {Independent Researcher, West Bengal, India}
\affil[3]{Department of Computer Science \& Engineering, Indian Institute of Technology Kharagpur, West Bengal, India}
\affil[4]{Cloud Computing and Distributed Systems (CLOUDS) Laboratory, School of Computing and Information Systems
The University of Melbourne, Australia}

\corraddress{Riya Samanta PhD, Department of Computer Science \& Engineering, Techno India University, West Bengal, India}
\corremail{riya.s@technoindiaeducation.com}

\runningauthor{Riya et al.}

\begin{document}

\begin{frontmatter}
\maketitle

\begin{abstract}
Modern LLM-based applications increasingly require multi-stage execution, persistent intermediate state, retry semantics, and auditable usage accounting. However, many LLM applications are still implemented as stateless prompt- \\ response wrappers or session-bounded conversational systems, which makes them difficult to recover, audit, and reproduce after interruption or failure. We propose \textbf{AgentR}, a stateful architecture for LLM workflow systems that enables persistence and recovery, instantiated through scientific literature review as a representative use case. AgentR represents research intent, generated queries, candidate-paper assessments and gap analyses as durable workflow artifacts, and executes the pipeline through asynchronous BullMQ workers backed by Redis, with PostgreSQL as the persistence store. The design includes explicit processing state transitions, retries with exponential backoff, orphan job detection, credit-aware pre-checks, ACID token-cost logging, and Type-2 slowly changing pricing records. We evaluate AgentR on telemetry collected from a prototype deployment. At the LLM stage, the system achieves 99.2\% job completion, and mean latencies of 9.0~s, 18.9~s, and 25.4~s for intent decomposition, query generation, and paper scoring, respectively. Parallel scoring allows for analytical latency modeling from observed calls, leading to as much as 4.3$\times$ wall-clock speedup over sequential execution. The results provide preliminary proof-of-concept that persistent state machine design, asynchronous orchestration, and cost-aware usage logging can enable improved observability, recoverability, and operational accountability in LLM workflow systems. The prototype implementation of AgentR is publicly available at: \url{https://github.com/RiyaSamanta/AgentR-public}.

\end{abstract}
\end{frontmatter}

\section{Introduction}

Scientific literature reviews are inherently iterative, multi-session activities. Researchers read papers in the morning, tweak their search strategies after lunch, and re-evaluate earlier assessments in the evening. They build up understanding over days, or weeks, or sometimes months. Teaching, meetings, or the need to read background material can break up a session of literature review. The context of reasoning, what was searched, what papers were reviewed and what gaps were found must be in place and not lost or unlinked when the researcher comes back. The need for \emph{persistent state} between sessions is fundamental to researchers' productivity and quality of research but is largely unmet by existing LLM-based literature review tools.

In addition to session persistence, the literature review has an inherent \emph{workflow structure}. In general, a good researcher should: (a) decompose a research problem into constituent axes such as methodological approach, application domain, constraints, and contribution type; (b) map this structured understanding to various search strategies; (c) assess retrieved papers along multiple analytical dimensions such as direct competition, supporting relevance, and gap identification; and (d) synthesize these assessments into an actionable research landscape~\citep{macfarlane2022search,wohlin2022successful}. These stages are not independent: intermediate decisions constrain downstream analysis, and the full reasoning trace needs to be tracked for reproducibility and audit.

The recent explosion of LLM-based literature tools has democratized the preliminary exploration of literature. Tools like Semantic Scholar~\citep{semanticscholar}, Scholarcy~\citep{scholarcy}, and SciSpace~\citep{scispace} are paper summarization systems that help users distill individual papers into summaries, key claims, or structured reading aids. Retrieval-oriented research assistants such as Consensus~\citep{consensus}, Semantic Scholar, and SciSpace support natural-language paper search, evidence discovery, and information extraction over large scholarly corpora. ResearchRabbit~\citep{researchrabbit2026} and Connected Papers~\citep{connectedpapers2026} allow literature discovery by mapping papers based on citation and similarity. Researchers also extensively use general-purpose chatbot systems such as ChatGPT~\citep{chatgpt}, Claude~\citep{claude}, Gemini~\citep{gemini}, and Perplexity~\citep{perplexity} to upload papers, ask follow-up questions, compare contributions, and draft literature-review text.

However, these tools are mostly built for document-level summarization, single-query retrieval, question answering, or session-bounded dialogue. They do not inherently model literature review as an ongoing executable workflow with explicit research-intent state, generated-query artifacts, paper-level analytical records, lifecycle transitions, recovery semantics, and auditable cost traces. This architectural gap motivates AgentR, which treats literature review as a durable, multi-stage LLM workflow rather than a sequence of isolated LLM interactions.

From a software-systems perspective, AgentR is evaluated as a quality-aware architecture for LLM workflow systems. The study focuses on reliability, recoverability, traceability, auditability, observability, and cost accountability as first-class software-quality attributes. Literature review is used as the case-study domain because it requires long-running, multi-stage, interruption-tolerant reasoning workflows; however, the architectural patterns are applicable to other LLM-enabled software services requiring persistent state and auditable execution.

\subsection{Problem Gap}
Preliminary literature exploration tools, such as recent LLM-based literature review tools, have made literature analysis easier through summarization, semantic retrieval, citation mapping, and conversational analysis. However, most existing systems still treat literature review as a series of isolated interactions rather than a sustained research workflow. An LLM takes as input an abstract, problem statement, or snippet of a paper and provides a one-shot response (direct prompting). While such outputs can enable rapid interpretation, they do not result in structured research intent, coordinated search strategies, or reusable intermediate artifacts.

Retrieval-augmented generation (RAG) grounds LLM responses on retrieved documents to improve factual accuracy~\citep{lewis2020rag}. Nevertheless, standard RAG pipelines follow a `query-retrieve-generate pattern'. They normally do not decompose a research problem into the methodological, application, constraint, and contribution dimensions, nor do they keep each analytical stage as durable state for subsequent reasoning. Likewise, systems like ChatGPT interactively offer session-level context, but this context is conversational rather than workflow-oriented. It is not maintained as a structured, queryable representation of research intent.

Recent agentic frameworks such as AutoGen~\citep{wu2023autogen}, MetaGPT~\citep{hong2023metagpt}, LangGraph~\citep{langgraph2026}, and CrewAI~\citep{crewai2026} extend LLM applications toward multi-agent collaboration and multi-stage orchestration. Some contemporary frameworks now support persisted execution state and workflow resumption. The remaining gap addressed here is therefore not persistence in isolation, but an integrated scientific-workflow architecture that couples durable research artifacts and explicit lifecycle semantics with recovery-aware execution, structured LLM-output validation, and auditable token/cost accounting.

Thus, the central gap is not the absence of another summarization, retrieval, or chatbot interface, but the absence of a software architectural model that formalizes LLM-assisted analysis as a persistent, stateful, recoverable, and auditable workflow. AgentR addresses this gap by treating literature review as a multi-stage agentic process in which research intent, generated queries, retrieved candidates, paper-level judgments, gap analyses, and synthesis outputs are stored as durable artifacts with lifecycle-aware orchestration and recovery-aware execution.

\subsection{Contributions}
We propose AgentR, a persistent, recovery-aware, and cost-auditable software architecture for LLM workflow systems. The architecture is instantiated through a literature-intelligence pipeline that transforms literature review from isolated LLM interactions into a durable multi-stage workflow with structured state, asynchronous execution, and accountable LLM usage. AgentR is not positioned as a paper summarizer, a RAG retrieval system, or a prompt-engineering wrapper; rather, it is a prototype system for studying persistency, recoverability, traceability, and cost accountability in stateful LLM workflows. 

This paper makes the following contributions:

\begin{enumerate}[leftmargin=*]
\item A \textbf{stateful software architecture for auditable LLM workflow systems} that represents research intent, generated queries, paper assessments, and gap-oriented analyses as durable, queryable artifacts rather than transient prompt responses.

\item A \textbf{structured literature-intelligence pipeline} that decomposes a research problem into scientific intent, generates multi-strategy search queries, and evaluates candidate papers along semantic similarity, direct-competitor relevance, supporting-work relevance, and gap-assessment dimensions.

\item A \textbf{recovery-aware asynchronous execution model} with implemented Redis-level retry, explicit processing states, exponential backoff, orphan-job detection, insufficient-credit handling, and a partially implemented database-reconstruction path for Redis-state-loss recovery.

\item A \textbf{cost-aware execution and audit model} that records token usage, model pricing, latency, and credit deductions using persistent usage logs and transaction-safe balance updates.

\item A \textbf{prototype evaluation} using development telemetry to characterize job completion, stage-wise latency, token cost, analytically derived parallelism speedup, and scoring-stability estimates.

\end{enumerate}

\subsection{Research Questions}
\label{sec:research_questions}

To align the evaluation with the software-systems focus of this work, we study AgentR through the following research questions:

\begin{enumerate}[leftmargin=*]
    \item \textbf{RQ1:} To what extent can a persistent state-machine architecture support completion and observability of multi-stage LLM workflows?
    \item \textbf{RQ2:} How does asynchronous queue-based execution affect end-to-end latency compared with sequential LLM execution?
    \item \textbf{RQ3:} What token and cost-accounting characteristics are observed when LLM workflow execution is logged at per-call granularity?
    \item \textbf{RQ4:} What recoverability mechanisms are supported by the implemented retry, orphan-detection, insufficient-credit handling, and resumption paths?
\end{enumerate}

The rest of this paper is organized as follows. Section~\ref{sec:related} is the detailed literature review. Section~\ref{sec:overview} presents the design goals, layered architecture, implementation scope, and end-to-end pipeline of AgentR. Section~\ref{sec:state} describes the persistent research-state model and processing-state semantics. Section~\ref{sec:pipeline} details the multi-stage literature-intelligence pipeline, including intent decomposition, query generation, and paper scoring. Section~\ref{sec:orchestration} explains the asynchronous orchestration model, while Section~\ref{sec:scoring} presents the scientific scoring framework. Section~\ref{sec:credit} discusses cost-aware execution and audit logging, and Section~\ref{sec:recovery} describes failure classification and resumable execution. Section~\ref{sec:experiments} reports the prototype evaluation.  Section~\ref{sec:conclusion} concludes the paper, and Section~\ref{sec:future} outlines future work.

\begin{table}[!t]
\caption{Feature-oriented comparison of representative related approaches.}
\label{tab:related_comparison}
\centering
\footnotesize
\setlength{\tabcolsep}{3.0pt}
\renewcommand{\arraystretch}{1.18}
\begin{tabular}{@{}p{3.25cm}*{7}{>{\centering\arraybackslash}p{1.32cm}}@{}}
\toprule
\textbf{Representative work / system} &
\textbf{\shortstack{Literature\\analysis}} &
\textbf{\shortstack{Multi-stage\\orchestration}} &
\textbf{\shortstack{Durable\\workflow state}} &
\textbf{\shortstack{Recovery /\\resume}} &
\textbf{\shortstack{LLM output\\validation}} &
\textbf{\shortstack{Gap-oriented\\analysis}} &
\textbf{\shortstack{Per-call\\cost audit}} \\
\midrule
Semantic Scholar, SciSpace, Scholarcy~\citep{semanticscholar,scispace,scholarcy} & $\checkmark$ & -- & -- & -- & -- & -- & -- \\
Consensus and Scite~\citep{consensus,scite2026} & $\checkmark$ & -- & -- & -- & -- & Partial & -- \\
ResearchRabbit and Connected Papers~\citep{researchrabbit2026,connectedpapers2026} & $\checkmark$ & -- & -- & -- & -- & Partial & -- \\
AutoGen and MetaGPT~\citep{wu2023autogen,hong2023metagpt} & -- & $\checkmark$ & Partial & Partial & Partial & -- & -- \\
LangGraph and CrewAI~\citep{langgraph2026,crewai2026} & -- & $\checkmark$ & $\checkmark$ & $\checkmark$ & Partial & -- & -- \\
Airflow, Prefect, Snakemake~\citep{airflow2026,prefect2026,molder2021snakemake} & -- & $\checkmark$ & $\checkmark$ & $\checkmark$ & -- & -- & -- \\
Galaxy and Taverna~\citep{afgan2018galaxy,wolstencroft2013taverna} & Partial & $\checkmark$ & $\checkmark$ & Partial & -- & -- & -- \\
Forecite and GapGPT~\citep{bai2023forecite,shamshoum2024gapgpt} & $\checkmark$ & -- & -- & -- & -- & $\checkmark$ & -- \\
\textbf{AgentR (this work)} & $\checkmark$ & $\checkmark$ & $\checkmark$ & $\checkmark$ & $\checkmark$ & $\checkmark$ & $\checkmark$ \\
\bottomrule
\end{tabular}
\end{table}

\section{Related Work}
\label{sec:related}

\subsection{LLM-Based Literature Analysis Tools}

Applying large language models to scientific literature has led to a range of tools. Semantic Scholar’s TLDR feature~\citep{cohan2020specter,semanticscholar} generates single-paper abstracts with fine-tuned models. Consensus~\citep{consensus} allows for natural-language question answering on paper databases, typically using retrieval-augmented generation (RAG) pipelines where a user query is embedded, similar documents are retrieved, and an LLM generates a response conditioned on the retrieved context~\citep{lewis2020rag}. Scite~\citep{scite2026} provides citation classification and smart citations to show if a cited paper provides supporting or contrasting evidence. Semantic Scholar employs AI to classify citation intent and tailor research feeds~\citep{ammar2018construction}. Document-level summarization and key-fact extraction are provided by SciSpace~\citep{scispace} and Scholarcy~\citep{scholarcy}.

The literature-analysis tools considered here primarily emphasize document-level summarization, retrieval, citation assistance, or conversational exploration. Their user-facing workflows do not generally expose the literature-review process as a durable state machine in which structured research intent, generated queries, paper-level assessments, and failure states are persisted as first-class workflow artifacts. This distinction is important for long-running review tasks in which intermediate analytical outputs must remain queryable and reusable across sessions.

\subsection{Agentic LLM Systems}

Recent work has explored multi-agent and agentic architectures for complex reasoning tasks. AutoGen~\citep{wu2023autogen} supports multi-agent conversations in which specialized agents collaborate through message passing, while MetaGPT~\citep{hong2023metagpt} applies structured agent roles to software-engineering workflows. ReAct~\citep{yao2023react} formalized the interleaving of reasoning and tool-use actions, and Toolformer~\citep{schick2023toolformer} demonstrated how language models can learn to invoke external tools. More recent orchestration frameworks such as LangGraph~\citep{langgraph2026} and CrewAI~\citep{crewai2026} go further by supporting stateful multi-step workflows; their current frameworks include mechanisms for persistence and resumption of long-running executions.

These developments mean that persistence and recovery are no longer unique capabilities of a single agent framework. The architectural distinction pursued by AgentR is instead the integration of those orchestration mechanisms with \emph{domain-level durable research artifacts}, explicit literature-workflow lifecycle states, LLM-output validation, paper-level gap assessments, and token-level cost accounting. In other words, AgentR is positioned as a software architecture for auditable scientific LLM workflows rather than as a general-purpose agent runtime.

\subsection{Scientific Workflow Systems}

Workflow management systems such as Apache Airflow~\citep{airflow2026}, Prefect~\citep{prefect2026}, and Snakemake~\citep{molder2021snakemake} leverage DAG-based task orchestration with retry semantics and state persistence. Galaxy~\citep{afgan2018galaxy} and Taverna~\citep{wolstencroft2013taverna} mainly target scientific workflows. These systems excel at orchestrating deterministic, compute-bound pipelines but lack native support for LLM-specific concerns like non-deterministic outputs needing structured validation, token-level cost tracking, and prompt-template versioning.

AgentR covers this gap by adding workflow orchestration semantics (queue-driven execution, state machines, retry with backoff) with LLM-oriented infrastructure (structured output validation via Zod schemas, JSON-mode enforcement, temperature-controlled generation, and per-call cost accounting). Existing workflow systems provide durable orchestration, and existing LLM literature tools provide retrieval or summarization, but AgentR combines persistent research-state modeling, LLM-specific structured validation, paper-level analytical artifacts, and token-level financial accounting in a single literature-intelligence workflow. The design draws on established patterns from workflow management~\citep{deelman2009workflows} and applies them to the specific challenges of LLM-driven scientific reasoning.

\subsection{Research Gap Analysis}
The automated identification of research gaps has been explored with bibliometric methods~\citep{small1973cocitation}, citation network analysis~\citep{klavans2009consensus}, and more recently LLM-based approaches. Forecite~\citep{bai2023forecite} uses citation graphs to discover underrepresented research topics. GapGPT~\citep{shamshoum2024gapgpt} uses LLMs to identify gaps in specific collections of papers. These techniques are generally one-shot analyses, and do not use the structured, pairwise comparison approach of AgentR, in which each candidate paper is evaluated relative to the researcher’s specific intention along multiple analytical dimensions, producing persistent, queryable gap evaluations.

\subsection{Structured Information Extraction from Scientific Text}

The extraction of structured information from scientific papers is a long-standing NLP task, from rule-based systems~\citep{teufel2009annotation} to neural approaches~\citep{luan2018multitask}. Recent work uses LLMs to extract claims~\citep{wadden2020fact}, methods~\citep{dunn2022extracting}, and contributions~\citep{chan2023scirepeval} from scientific text. AgentR extends this tradition by extracting a structured representation of \emph{research intent}, including problems, methodologies, domains, constraints, contribution types, and seed keywords. These pillars are the durable state drivers for all downstream stages of the scientific research workflow pipeline.

\subsection{Comparative Synthesis of Related Approaches}

Table~\ref{tab:related_comparison} summarizes the capabilities emphasized by representative systems discussed above. The comparison is feature-oriented rather than a ranking: a check mark denotes explicit support, ``Partial'' denotes related functionality that is narrower or requires additional configuration, and ``--'' denotes a capability that is not a primary feature of the cited system. Importantly, modern agent frameworks already provide substantial workflow-state and recovery support. The differentiating point for AgentR is therefore the \emph{combination} of workflow durability with research-specific persistent artifacts, LLM-output validation, gap-oriented paper analysis, and per-call token/cost auditability.

\section{System Overview}
\label{sec:overview}

\subsection{Design Goals}

AgentR is built around five requirements that are derived from the shortcomings of existing systems:

\begin{itemize}[leftmargin=*]
    \item \textbf{Persistent Research State.} The literature analysis must produce lasting, structured artifacts that can survive session termination, and can be queried by downstream stages without re-computation.
    \item \textbf{Explicit Workflow Lifecycle.} Pipeline stages should have observable, auditable, and recoverable, well-defined state transitions (not-started, in-progress, completed, failed).
     \item \textbf{Asynchronous Execution.}  LLM calls are inherently slow, taking from seconds to minutes. The system should not block the user interaction while it is processing but processing should be asynchronous with status visibility.
    \item \textbf{Resilience to Failure.} Rate limits, network issues and running out of credits can cause the LLM API calls to fail. The system should distinguish between transient and permanent failures and retry as appropriate. The system should permit manual resumption following recoverable failures.
    \item \textbf{Accountability for Finance.} Every call to an LLM has a measurable cost. The system must track costs on a per-token basis, maintain auditable financial records and prevent operations if funds are insufficient.
\end{itemize}

\begin{figure}[!t]
    \centering
    \includegraphics[width=0.98\linewidth,height=\textheight,keepaspectratio]{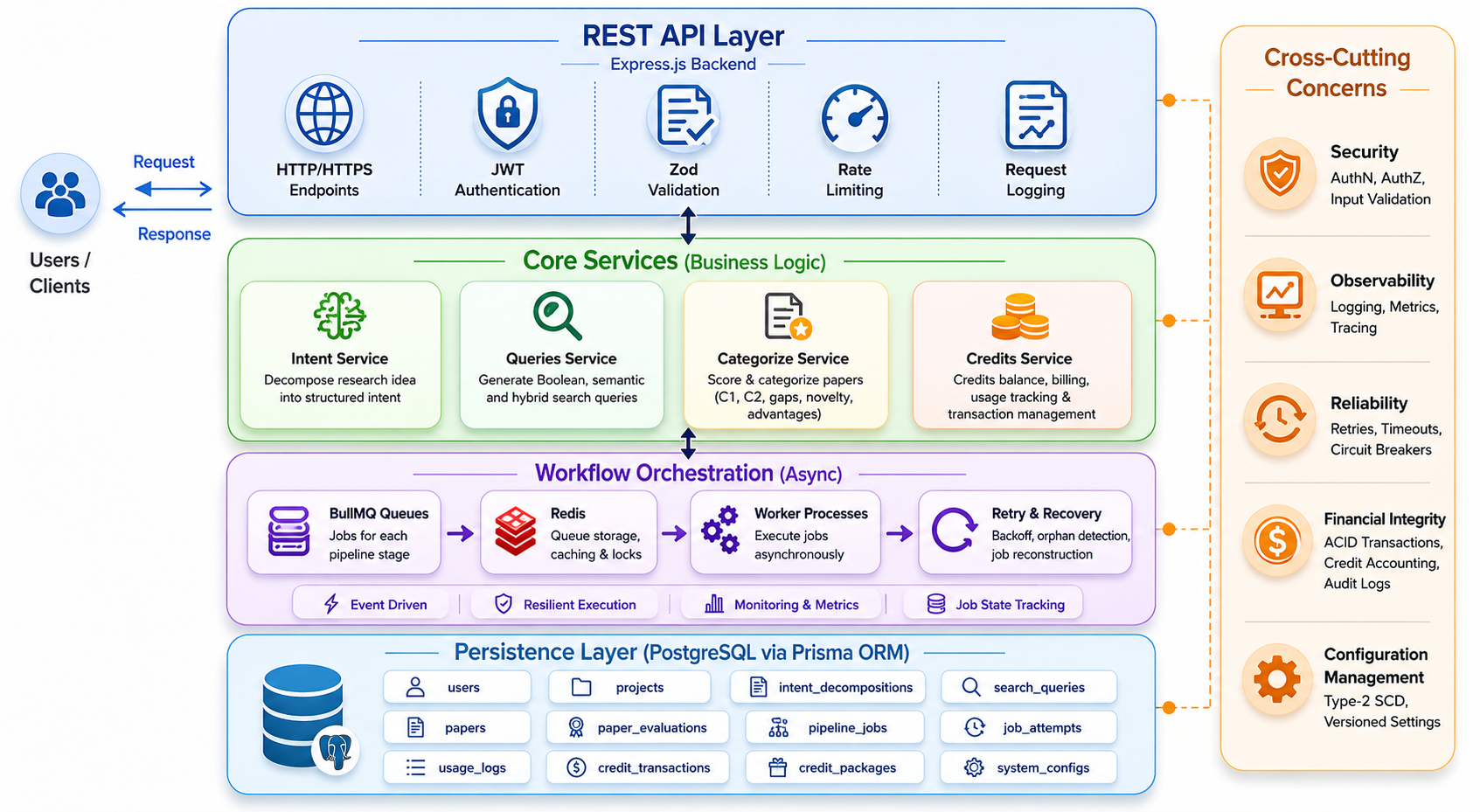}
    \caption{Layered architecture of the AgentR framework}
    \label{fig:architecture}
\end{figure}
\subsection{Architecture}

We design the AgentR architecture, which consists of four layers as depicted in Fig.~\ref{fig:architecture}:

\begin{enumerate}[leftmargin=*]
    \item \textbf{API Layer.}  A RESTful Express.js server that manages authentication (JSON Web Tokens and refresh tokens), request validation (Zod schemas), rate limiting, and consistent response formatting.
    \item \textbf{Service Layer.} Modules of business logic for each pipeline stage: intent decomposition, query generation, paper scoring, and auxiliary services (credits, usage logging, model pricing, payments).
    \item \textbf{Orchestration Layer.}  Three BullMQ queues based on Redis for asynchronous job dispatch, retry with exponential backoff, worker lifecycle. Each queue handles a specific pipeline domain: project initiation, paper grading, and email notification.
    \item \textbf{Persistence Layer.} A PostgreSQL database managed with Prisma Object-Relational Mapping (ORM), containing 13 tables to store research projects, candidate papers, LLM usage, financial transactions, configuration history, and job state.
\end{enumerate}

\subsection{Implementation Scope}
AgentR is a Node.js backend with 7,298 lines of TypeScript in 83 source files, 13 Prisma data models, 4 Prisma enums, and 63 documented REST API endpoints. 101 React TypeScript files make up the frontend. The system comprises 7 Zod validation schema files with 24 named schemas, 3 LLM prompt template files and 3 BullMQ queues, backed by Redis. The complete set of system statistics is in Table~\ref{tab:systemstats}.

\begin{table}[htbp]
\caption{AgentR System Implementation Statistics}
\label{tab:systemstats}
\centering
\begin{tabular}{@{}lr@{}}
\toprule
\textbf{Component} & \textbf{Count} \\
\midrule
Backend TypeScript files & 83 \\
Frontend TS/TSX files & 101 \\
API route files & 16 \\
Controller files & 13 \\
Service files & 24 \\
Worker files & 4 \\
Middleware files & 9 \\
Zod validation schema files & 7 \\
LLM prompt template files & 3 \\
Prisma data models & 13 \\
Prisma enums & 4 \\
BullMQ background queues & 3 \\
Background job types & 4 \\
Backend TypeScript LOC & 7,298 \\
\bottomrule
\end{tabular}
\end{table}

\subsection{Pipeline Overview}

The AgentR pipeline executes a research project in three implemented stages, with further stages planned for automatic paper retrieval and cross-paper synthesis:

\textbf{Stage 1: Intent Decomposition.} The input to the system is a research abstract. Six structured fields are extracted. These are: problem statement, methodological approaches, application domains, constraints, contribution types, and seed keywords. This structured representation is the research state for all further stages.

\textbf{Stage 2: Query Generation.} Using the structured intent as input, the system generates three types of search artifacts: a Boolean query with AND/OR operators for database-specific search, an expanded keyword set (10-15 terms), and exactly 10 optimized search query strings for general-purpose academic search engines.

\textbf{Stage 3: Paper Scoring.} For each candidate paper, the system performs four concurrent analyses in a single LLM call: (a)~semantic similarity with dimensional overlap assessment; (b)~Category~1 (C1) direct-competitor evaluation with rubric-based scoring (0–10); (c)~Category~2 (C2) supporting-work evaluation with contribution-type classification; and (d)~research-gap identification with novelty and advantage analysis. These three analytical dimensions (semantic matching, categorization, and gap analysis) are merged into a single LLM call to reduce latency, cost, and inter-call inconsistency (see Section~\ref{sec:scoring}).

\begin{figure}[!t]
    \centering
\includegraphics[width=\linewidth,height=0.8\textheight]{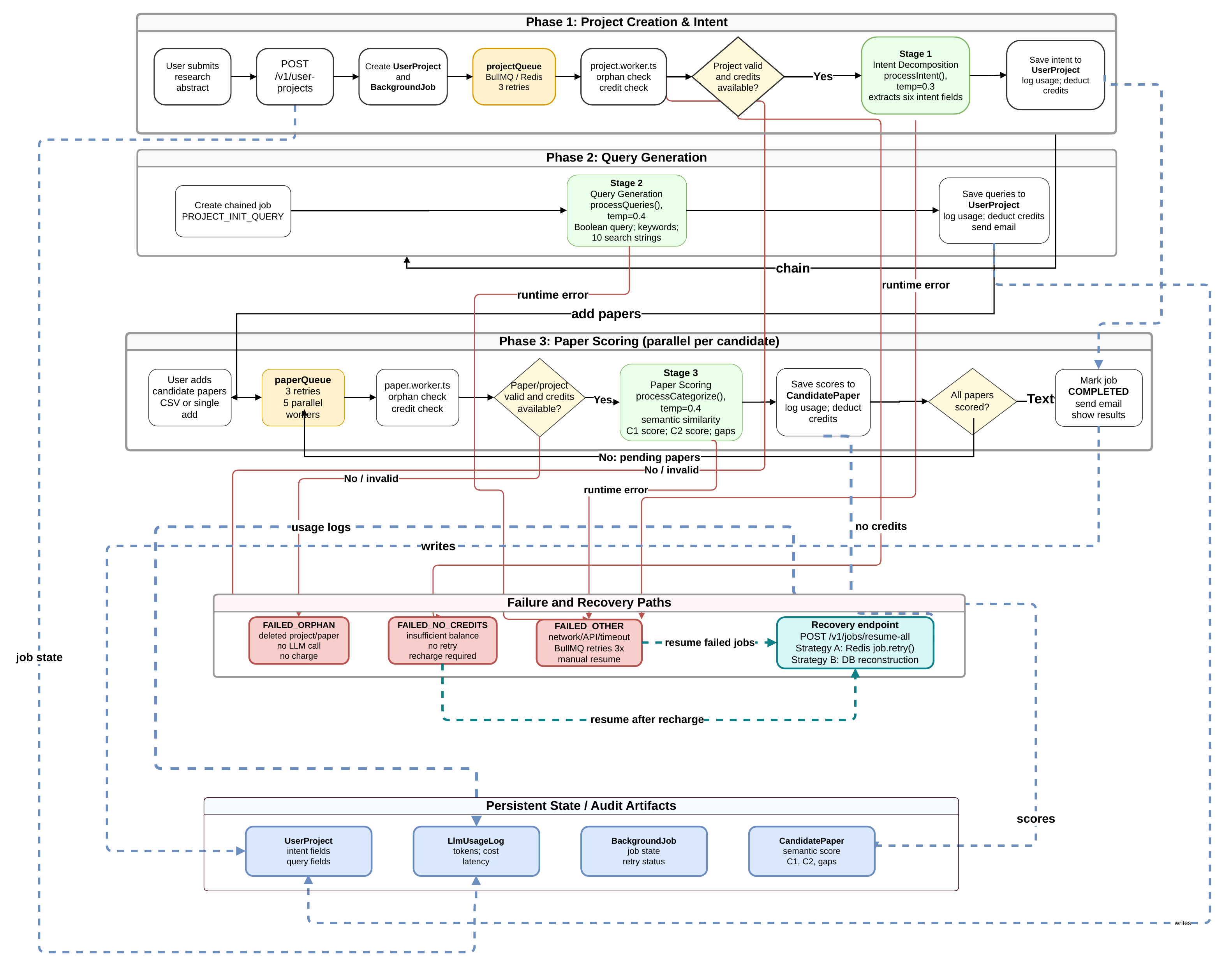}
    \caption{End-to-end activity diagram of the AgentR pipeline.}
    \label{fig:end_to_end_pipeline}
\end{figure}

Fig.~\ref{fig:end_to_end_pipeline} presents the end-to-end activity flow of AgentR from project creation to intent extraction, query generation, parallel candidate-paper scoring, persistence, credit accounting, and failure recovery. The diagram shows that AgentR is not a simple prompt-response interaction. Instead, each analysis step is an asynchronous job with explicit validation, durable persistence, and resumable failure handling. Additionally, Table~\ref{tab:quality_attributes} maps AgentR's architectural mechanisms and the quality attributes evaluated in this paper.

\begin{table}[htbp]
\caption{AgentR Architectural Mechanisms and Supported Software-Quality Attributes}
\label{tab:quality_attributes}
\centering
\begin{tabular}{p{3.2cm}p{5.2cm}p{4.2cm}}
\hline
\textbf{Quality attribute} & \textbf{AgentR mechanism} & \textbf{Evidence in evaluation} \\
\hline
Reliability & BullMQ retry, exponential backoff, worker isolation & Job completion rate, failure statistics \\
Recoverability & ProcessingStatus state machine, resume endpoint, Redis retry & Recovery architecture and resumption semantics \\
Traceability & Durable project, paper, job, and usage records & Persistent state and audit logs \\
Auditability & LLM usage logs, pricing history, credit accounting & Token and cost efficiency analysis \\
Maintainability & Layered architecture, typed schemas, service separation & Implementation statistics and modular design \\
Observability & Queryable job states and background-job records & Workflow status tracking \\
\hline
\end{tabular}
\end{table}

\section{Persistent Research-State Modeling}
\label{sec:state}

\subsection{State Representation}

AgentR models a research project as a persistent entity with multiple typed state fields. The \texttt{UserProject} table stores both the original input (the researcher's abstract) and all pipeline outputs, including the Stage~1 intent decomposition fields (\texttt{problemStatement}, \texttt{methodologies}, \texttt{applicationDomains}, \texttt{constraints}, \texttt{contributionTypes}, \texttt{keywordsSeed}) and Stage~2 query generation outputs (\texttt{booleanQuery}, \texttt{expandedKeywords}, \texttt{searchQueries}).

Each candidate paper is similarly represented as a persistent entity with its full analytical output: semantic similarity score $s \in [0,1]$; four dimensional overlap assessments $o_i \in \{\text{none}, \text{low}, \text{medium}, \text{high}\}$ for $i \in \{\text{problem}, \text{method}, \\ \text{domain}, \text{constraint}\}$; C1 and C2 scores with justifications, strengths, and weaknesses; and research gap artifacts (\texttt{researchGaps}, \texttt{userNovelty}, \texttt{candidateAdvantage}).

\subsection{State Machine Semantics}

\begin{figure}[!t]
    \centering
    \includegraphics[width=0.88\linewidth,height=0.50\textheight,keepaspectratio]{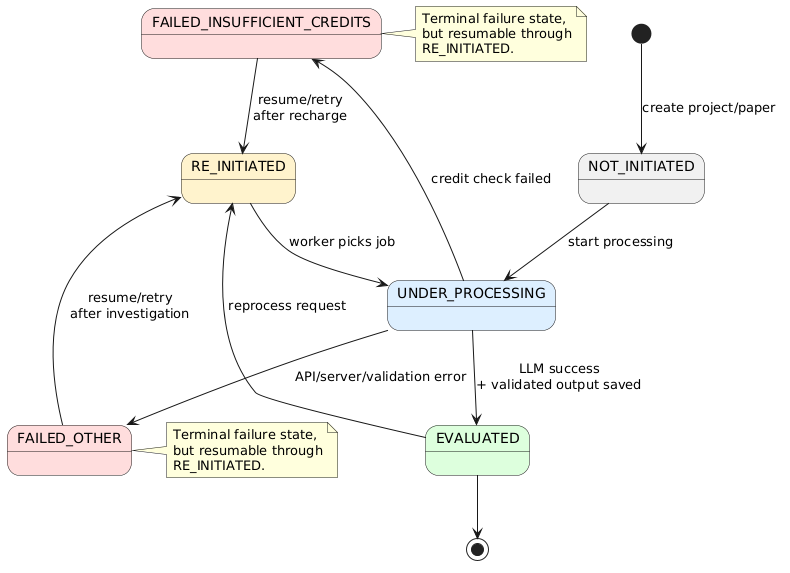}
    \caption{Processing status state machine.}
    \label{fig:statemachine}
\end{figure}

The \texttt{ProcessingStatus} enum defines a six-state finite automaton shared across projects and papers, as shown in Fig.~\ref{fig:statemachine}.

This state machine has several properties that distinguish it from the implicit state in conversational systems:

\begin{enumerate}[leftmargin=*]
    \item \textbf{Observable.} Any client can query the current state of any pipeline stage at any time.
    \item \textbf{Durable.} State is persisted in PostgreSQL, not in worker memory or conversation context.
    \item \textbf{Recoverable.} Terminal failure states are resumable via the \texttt{RE\_INITIATED} transition.
    \item \textbf{Auditable.} State transitions are timestamped and traceable through the \texttt{BackgroundJob} table.
\end{enumerate}

\subsection{Soft Delete and Cascade Semantics}

Project deletion follows a cascading soft-delete pattern, implemented as a Prisma database transaction: deleting a project atomically sets \texttt{isDeleted=true} on the project, all its candidate papers, and all associated LLM usage logs. This keeps audit trails intact and logically removes the project from user queries. The workers check these soft-delete flags before expensive LLM calls to avoid wasted computation on deleted resources.

\section{Multi-Stage Scientific Reasoning Pipeline}
\label{sec:pipeline}

\subsection{Stage 1: Structured Intent Decomposition}

The intent decomposition stage transforms a free-text research abstract into a structured representation of research intent. Given an abstract $a$, the intent decomposition function $\mathcal{I}$ produces a tuple:
\begin{equation}
    \mathcal{I}(a) = \langle p, M, D, C, T, K \rangle
    \label{eq:intent}
\end{equation}
where $p$ is the problem statement, $M$ is the set of methodological approaches, $D$ is the set of application domains, $C$ is the set of constraints, $T$ is the set of contribution types, and $K$ is the set of seed keywords.

The LLM is prompted with a system instruction specifying the six extraction targets and a user prompt containing the abstract. Temperature is set to 0.3 for deterministic extraction. The response is enforced to be valid JSON via OpenAI's \texttt{json\_object} response format and validated against a Zod schema before persistence. Any validation failure triggers a retry via the BullMQ backoff mechanism.

Unlike free-text summarization, structured intent decomposition generates typed and indexed fields for downstream stages to directly consume. The query generator feeds on the problem statement, the scoring rubric on the methodological axes, the overlap analysis on the constraint set.
\subsection{Stage 2: Multi-Strategy Query Generation}

The query generation stage translates structured intent into diverse search strategies. The input is the full intent tuple $\langle p, M, D, C, T, K \rangle$ from Stage~1. The query generator produces three artifact types:

\begin{enumerate}[leftmargin=*]
    \item \textbf{Boolean Query.} A structured query using AND/OR operators compatible with database-specific search engines (IEEE Xplore, Semantic Scholar,ACM Digital Library, PubMed).
    \item \textbf{Expanded Keywords.} 10--15 terms including synonyms, hyponyms, and related concepts not present in the original abstract.
    \item \textbf{Search Queries.} Exactly 10 distinct query strings optimized for general-purpose academic search engines, generated with explicit variation strategies: broad, narrow, method-specific, problem-specific, and domain-specific.
\end{enumerate}

Stage~2 uses temperature 0.4, which is slightly higher than Stage~1, to encourage diverse query formulations while maintaining relevance.

\subsection{Stage 3: Multi-Dimensional Paper Scoring}

The analytical core of the pipeline is the paper scoring. It produces a holistic evaluation in four dimensions in one LLM call per candidate paper.

\subsubsection{Semantic Similarity Assessment}

The scoring function computes an overall semantic similarity score $s \in [0, 1]$ and four dimensional overlap assessments $o_i \in \{\text{none}, \text{low}, \text{medium}, \text{high}\}$ for $i \in \{\text{problem}, \text{method}, \text{domain}, \text{constraint}\}$.

\subsubsection{Category 1: Direct Competitor Assessment}

The C1 score $c_1 \in [0, 10]$ evaluates whether a candidate paper addresses the same problem using a similar methodology. The scoring follows an explicit rubric embedded in the prompt, as shown in Table~\ref{tab:c1rubric}.

\begin{table}[htbp]
\caption{C1 Direct Competitor Scoring Rubric}
\label{tab:c1rubric}
\centering
\begin{tabular}{@{}cl@{}}
\toprule
\textbf{Score} & \textbf{Interpretation} \\
\midrule
9--10 & Near-identical problem and method \\
7--8 & Same problem, similar method with variations \\
5--6 & Same problem with different method, or vice versa \\
3--4 & Partial overlap in both dimensions \\
0--2 & Minimal overlap, not a competitor \\
\bottomrule
\end{tabular}
\end{table}

\subsubsection{Category 2: Supporting Work Assessment}

The C2 score $c_2 \in [0, 10]$ evaluates whether a candidate paper provides relevant supporting context. Each C2 assessment includes a contribution-type classification:

\[
\small
\begin{array}{l}
\texttt{type} \in \{ \texttt{methodology},\ \texttt{problem\_context},\\
\texttt{domain\_knowledge},\ \texttt{constraint\_analysis},\\
\texttt{related\_application},\ \texttt{theoretical\_foundation} \}
\end{array}
\]

\subsubsection{Research Gap Identification}

For each candidate paper, the system identifies three gap-related artifacts: (1)~\texttt{research\_gaps}: what the candidate paper does \emph{not} address that the researcher's work intends to (2)~\texttt{user\_novelty}: how the researcher's objective may lead to advancement beyond the candidate; and (3)~\texttt{candidate\_advantage}: what the candidate offers that the researcher's proposal does not.

\subsection{Merged Stage Design}

In Stage 3, the three analytical dimensions are merged into a single LLM call rather than three separate calls. This design choice leads to reduced latency (one round trip instead of three), reduced cost (one prompt prefix instead of three), and reduced inconsistency (the LLM evaluates all dimensions in a single reasoning context).  The tradeoff is a larger prompt and a more complex output schema, facilitated by Zod validation and JSON-mode enforcement.

\section{Asynchronous Workflow Orchestration}
\label{sec:orchestration}

AgentR uses three BullMQ queues, each with domain-specific retry semantics. Pipeline stages are connected through a chaining pattern where each worker, upon completing its stage, creates the next stage's job as a new entry in the appropriate queue:

\begin{enumerate}[leftmargin=*]
    \item Controller $\rightarrow$ \texttt{projectQueue(PROJECT\_INIT\_INTENT)}
    \item Worker completes Stage~1 $\rightarrow$ \texttt{projectQueue(PROJECT\_INIT\_QUERY)}
    \item Worker completes Stage~2 $\rightarrow$ \texttt{emailQueue(SEND\_EMAIL)}
\end{enumerate}

This pattern ensures that each stage has independent retry semantics, a failure in Stage~2 does not require re-executing Stage~1, and the pipeline is observable at each handoff point via the \texttt{BackgroundJob} table.

Before executing any expensive LLM call, each worker performs an orphan check: it queries the database to verify that the parent entity (project or paper) still exists and has not been soft-deleted. If the entity is missing or deleted: (1)~the job status is set to \texttt{FAILED\_ORPHAN}; (2)~the worker returns early without making any LLM call; (3)~no credits are deducted.

The job resume mechanism implements a dual-strategy recovery:

\textbf{Strategy A: Redis Retry.} If the BullMQ job object still exists in Redis (jobs are retained with \texttt{removeOnFail: false}), the system calls \texttt{job.retry()} to re-enqueue it directly.

\textbf{Strategy B: Database Reconstruction.} In the event that the BullMQ job has expired from Redis (e.g. as a result of a Redis restart), the system attempts to rebuild the job payload from durable database state by calling \texttt{reconstructJobPayload()}. This reconstruction path is only partially realized in the current prototype.


Controllers wrap all \texttt{queue.add()} calls in a 5-second timeout wrapper. If the timeout expires (e.g., Redis is unavailable), the \texttt{BackgroundJob} record is marked \texttt{FAILED}, but the API returns HTTP 202 Accepted. The user can retry the job later via the resume endpoint.

\begin{figure}[!t]
    \centering
    \includegraphics[width=\linewidth,height=0.6\textheight,keepaspectratio]{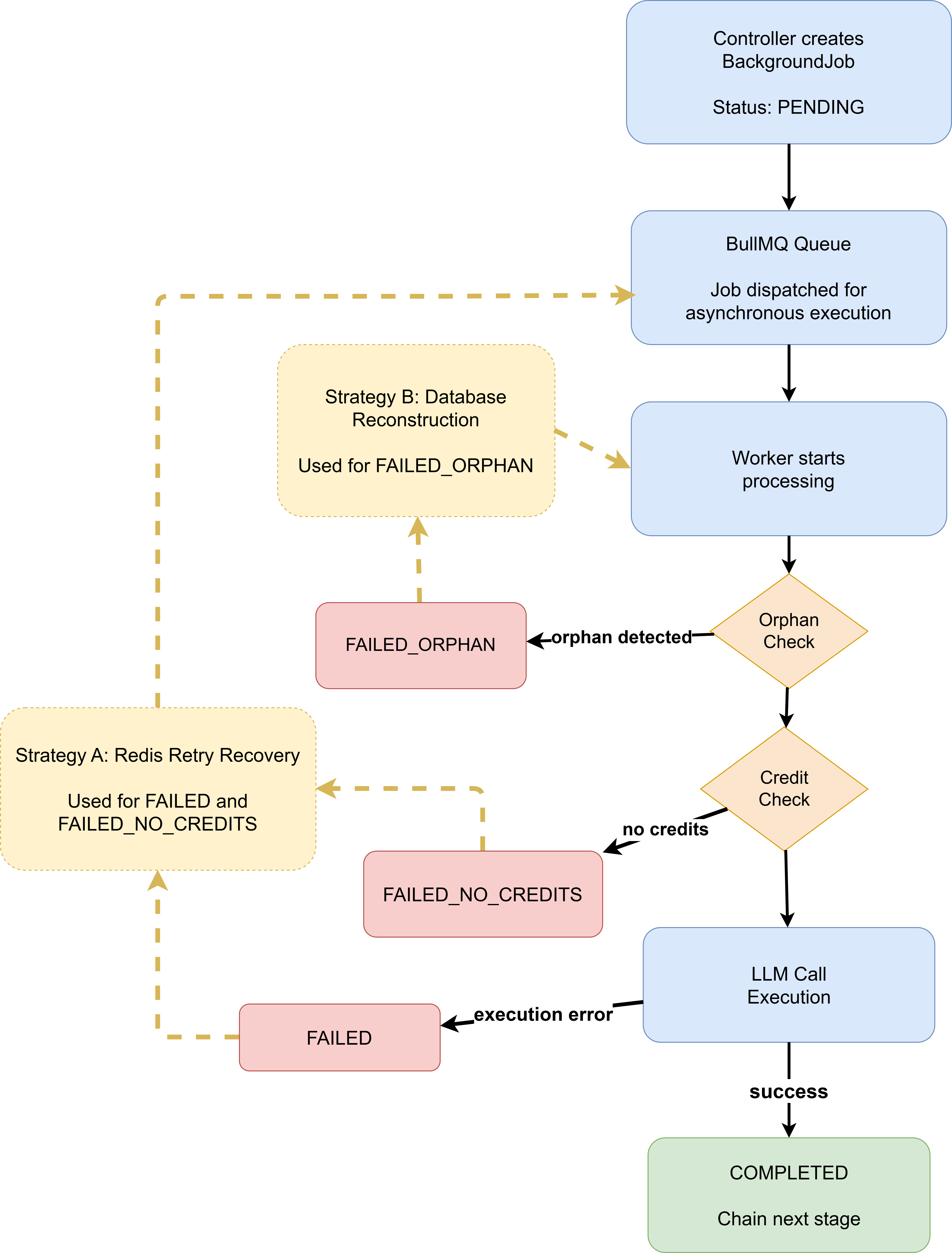}
    \caption{Job lifecycle and double-strategy recovery mechanism.}
    \label{fig:job_lifecycle_recovery}
\end{figure}
 \FloatBarrier

\section{Scientific Scoring Framework}
\label{sec:scoring}
The scoring rubrics for C1 and C2 are included in the LLM system prompt, with explicit definitions of score ranges. This design serves two purposes: (1) \emph{calibration} – by defining what each score range means, the rubric reduces inter-call variance in scoring; and (2) \emph{auditability} – the rubric is versioned along with the prompt template, allowing for reproducibility analysis.

OpenAI's \texttt{response\_format: \{type: ``json\_object''\}} mode forces all LLM outputs to be valid JSON. Outputs are validated upon receipt against Zod schemas for type correctness, range constraints, enum membership, and required fields. Validation failures are temporary errors and will trigger BullMQ retry.

Different temperature settings are used for different pipeline stages depending on their task characteristics, as shown in Table~\ref{tab:temperature}.

\begin{table}[htbp]
\caption{Temperature Settings by Pipeline Stage}
\label{tab:temperature}
\centering
\begin{tabular}{@{}lcc@{}}
\toprule
\textbf{Stage} & \textbf{Temperature} & \textbf{Rationale} \\
\midrule
Intent Decomposition & 0.3 & Deterministic extraction \\
Query Generation & 0.4 & Creative variation needed \\
Paper Scoring & 0.4 & Nuanced evaluation \\
\bottomrule
\end{tabular}
\end{table}

\section{Credit-Aware and Auditable Execution}
\label{sec:credit}

There are three tiers of the credit system:
\begin{enumerate}[leftmargin=*]
    \item \textbf{API Middleware Layer.}  A \texttt{checkCreditsMiddleware} intercepts incoming requests and returns HTTP 402 (Payment Required) if the user’s credit balance is zero or negative.
    \item \textbf{Pre-check Layer for Workers.} Each worker verifies that the user has at least a minimum credit balance before making an LLM call, a conservative threshold to prevent starting work that cannot be completed.
    \item \textbf{Post-execution Deduction Layer.}  \begin{enumerate} \item On a successful LLM call, the \texttt{logLlmUsage()} function runs an ACID transaction that atomically (a) creates a \texttt{LlmUsageLog} entry with token counts and USD cost, (b) reads the active USD-to-credits multiplier, (c) computes the credit deduction, and (d) decrements the user's balance. \end{enumerate}
    \end{enumerate}

Model pricing and system configuration are on a Type-2 slowly changing dimension (SCD) pattern. On a pricing change: (1)~set the \texttt{isLatest} flag of the current active record to \texttt{false} and its \texttt{effectiveTo} to the current timestamp; (2)~insert a new record with \texttt{isLatest = true}, \texttt{isActive = true} and \texttt{effectiveFrom = current\_timestamp}.

This stores the entire history of pricing changes which makes it possible to calculate the cost of historical LLM calls correctly and allows auditing and rolling back configuration changes to previous pricing. This pattern is implemented in three tables: \texttt{LlmModelPricing}, \texttt{CreditsMultiplierHistory}, and \texttt{DefaultCreditsHistory}.

The system has a two tier audit trail.  The \texttt{LlmUsageLog} table logs each LLM call with token counts, calculated USD cost, model, pipeline stage, latency, and denormalized project/paper names which persist after deletion of parent entity. This provides per-call cost traceability at the token level. The \texttt{UserCreditsTransaction} table keeps a full audit trail of all credit operations carried out by admins (signup allocation, manual recharge, manual deduction) with details of before and after balances and transaction type, enabling comprehensive admin credit management. The \texttt{MoneyTransaction} table stores payment gateway records. Credit deduction for LLM use is atomic with usage-logging in the same Prisma transaction, so that balance updates are ACID-consistent with usage records.

\begin{figure}[!t]
    \centering
\includegraphics[width=\linewidth,height=0.8\textheight,keepaspectratio]{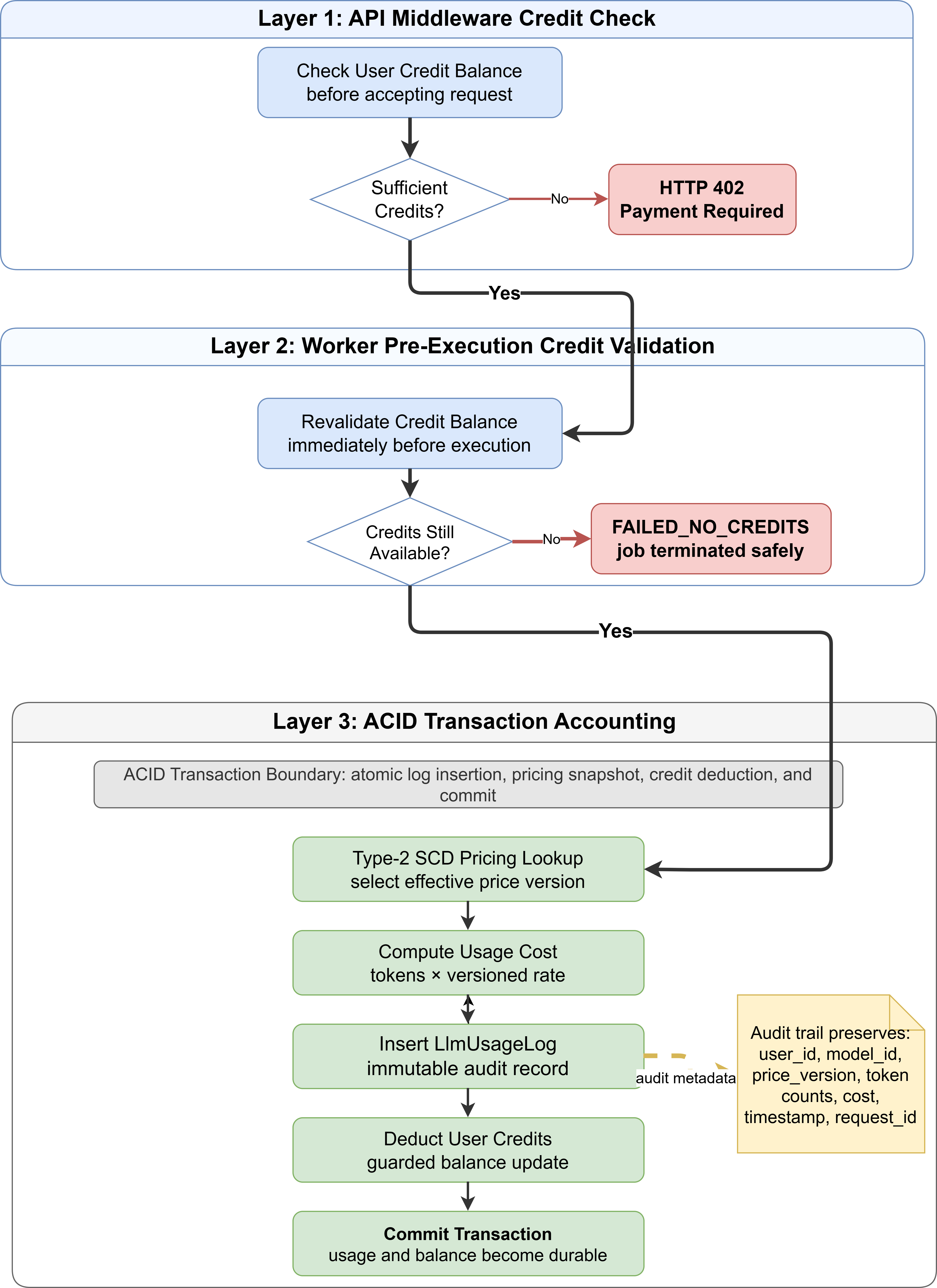}
    \caption{Three-layer credit architecture with ACID transaction guarantees.}
    \label{fig:creditflow}
\end{figure}

\section{Failure Recovery and Stateful Resumption}
\label{sec:recovery}

AgentR classifies failures into three categories with distinct handling strategies, as shown in Table~\ref{tab:failures}.

\begin{table}[htbp]
\caption{Failure Taxonomy and Handling Strategies}
\label{tab:failures}
\centering
\begin{tabular}{@{}lccc@{}}
\toprule
\textbf{Type} & \textbf{Retry} & \textbf{Resume} & \textbf{Credit} \\
\midrule
Transient & Yes (3$\times$) & Yes & No charge \\
Permanent\ (credits) & No & Yes & No charge \\
Permanent\ (orphan) & No & No & No charge \\
\bottomrule
\end{tabular}
\end{table}

Workers implement explicit error classification. For \texttt{INSUFFICIENT\_CREDITS} errors, the job is marked \\ \texttt{FAILED\_NO\_CREDITS} and the error is \emph{not} re-thrown to BullMQ, preventing wasted retries. For all other errors (network timeouts, API failures), the error \emph{is} re-thrown, allowing BullMQ to retry with exponential backoff.

The \texttt{POST /v1/jobs/resume-all} endpoint enables bulk recovery of all failed jobs for a user. For each failed job: (1)~orphan check: verify parent project exists; (2)~credit check: verify balance $> 0$; (3)~apply the available recovery path (Redis retry in the current implementation, with database reconstruction treated as a planned extension); (4)~transition job to \texttt{PENDING} and processing status to \texttt{RE\_INITIATED}.

\section{System Evaluation}
\label{sec:experiments}

We evaluate AgentR as a prototype implementation, focusing on the system architecture rather than domain-level superiority in literature-review quality. The evaluation is organized around the research questions defined in Section~\ref{sec:research_questions}: workflow completion and job status visibility address RQ1; latency and parallelism address RQ2; token and cost telemetry address RQ3; and retry, orphan-checking, insufficient-credit handling, and resumption semantics address RQ4.

\subsection{Evaluation Design}

\subsubsection{Dataset}

We analyze a development database snapshot containing 6 registered users, 14 projects (10 active, 4 soft-deleted), 44 candidate papers (43 active), 130 background jobs, and 72 LLM usage logs. The 10 active projects span 6 domain clusters: precision agriculture and IoT sensing (2 projects, 34 papers), TinyML and embedded machine learning (2 projects, 3 papers), medical imaging and edge AI (1 project, 3 papers), indoor navigation and autonomous systems (1 project, 2 papers), drone surveillance and remote sensing (1 project, 2 papers), and multimedia and video editing (1 project, 1 paper). Candidate papers were sourced from Semantic Scholar~\citep{ammar2018construction} and evaluated against their parent project's research abstract. Table~\ref{tab:corpus} summarizes the evaluation corpus.

\begin{table}[htbp]
\caption{Evaluation Corpus Summary}
\label{tab:corpus}
\centering
\begin{tabular}{@{}lr@{}}
\toprule
\textbf{Characteristic} & \textbf{Value} \\
\midrule
Registered users & 6 \\
Active projects (non-deleted) & 10 \\
Soft-deleted projects & 4 \\
Total candidate papers & 44 \\
Active candidate papers & 43 \\
Evaluated papers (Stage 3 complete) & 43 \\
Total background jobs & 130 \\
Total LLM usage logs & 72 \\
Projects with Stage 1+2 complete & 14/14 (100\%) \\
Domain clusters represented & 6 \\
\bottomrule
\end{tabular}
\end{table}

\subsubsection{Baselines}

We compare AgentR against analytically derived baselines representing different architectural approaches:

\begin{enumerate}[leftmargin=*]
    \item \textbf{AgentR-full.} The current system with 5-way parallel paper scoring, BullMQ retry with exponential backoff, orphan detection, and credit-aware execution.
    
    \item \textbf{GPT-Direct-Sequential.} The same three-stage prompts executed synchronously with no persistence, no retry, and no parallelism. If any stage fails, the entire pipeline must restart from Stage~1.
    \item \textbf{GPT-Direct-Plus-Retry.} Identical to GPT-Direct-Sequential but with a simple 3-retry loop on failure, preventing the comparison from being a strawman against a system with no error handling.
\end{enumerate}

 All baseline latency and cost estimates are derived analytically from the 72 observed LLM calls, using the same per-stage mean durations and token counts as AgentR. This ensures a fair comparison: the baselines differ only in architecture (parallelism, persistence, retry), not in prompt quality or model choice.

\subsection{Metrics}

\begin{itemize}[leftmargin=*]
    \item \textbf{Workflow Completion Rate (WCR):} Percentage of projects completing all pipeline stages (measured from database records).
    \item \textbf{Job Completion Rate (JCR):} Percentage of background jobs reaching COMPLETED status (measured from database records).
    \item \textbf{Per-Stage Latency:} LLM call duration per pipeline stage in milliseconds (measured from usage logs).
    \item \textbf{Parallelism Speedup:} Wall-clock latency ratio between parallel and sequential execution (analytically derived from measured per-stage latencies).
    \item \textbf{Token and Cost Efficiency:} Token consumption and USD cost per stage and per project (measured from usage logs; gpt-4o estimates analytically projected).
    \item \textbf{Exploratory Scoring Stability:} ICC and Cohen's $\kappa$ estimated through simulation-based test-retest modeling over observed single-pass score distributions; these values are not treated as live repeated-run reliability measurements.

\end{itemize}

\subsection{Results}

\subsubsection{Workflow Completion and Job Statistics}

We evaluated workflow completion on a dataset of 10 active research projects with 43 evaluated candidate papers, drawn from a development database backup. Table~\ref{tab:wcr} reports workflow completion rates.

\begin{table*}[htbp]
\caption{Workflow Completion Rate}
\label{tab:wcr}
\centering
\begin{tabular}{@{}lcccc@{}}
\toprule
\textbf{System} & \textbf{Active Projects} & \textbf{Stages 1+2 Complete} & \textbf{WCR (\%)} & \textbf{Failed Projects} \\
\midrule
AgentR & 10 & 10 & 100.0 & 0 \\
\bottomrule
\end{tabular}
\end{table*}

We note that this 100\% WCR reflects a controlled development dataset with near-zero failure rate. Production WCR would be lower. Table~\ref{tab:jobs} reports per-type job completion statistics from 130 observed background jobs.

\begin{table*}[htbp]
\caption{Job Completion and Failure Statistics}
\label{tab:jobs}
\centering
\footnotesize
\begin{tabular}{@{}lrrrrr@{}}
\toprule
\textbf{Job Type} & \textbf{\small Total} & \textbf{\small COMPLETED} & \textbf{\small FAILED} & \textbf{\small Completion (\%)} \\
\midrule
PROJECT\_INIT\_INTENT & 14 & 14 & 0 & 100.0 \\
PROJECT\_INIT\_QUERY & 14 & 14 & 0 & 100.0 \\
PAPER\_SCORING & 44 & 44 & 0 & 100.0 \\
SEND\_EMAIL & 58 & 57 & 1 & 98.3 \\
\midrule
\textbf{Total} & \textbf{130} & \textbf{129} & \textbf{1} & \textbf{99.2} \\
\bottomrule
\end{tabular}
\end{table*}

AgentR's job completion rate (99.2\%) is attributable to its retry mechanism and pre-flight credit verification. The single failure occurred in an email notification job.

\subsubsection{Recovery Architecture}

The source code and job history were examined to confirm the current recovery capabilities of AgentR. \textbf{Strategy A (BullMQ retry):}  The project and paper queues are configured to retry three times with exponential backoff (1,000~ms and 2,000~ms base delays, respectively). In case of transient failure (network timeout or API failure) the worker re-throws the error and BullMQ puts the job back in queue with an increasing delay. The retry strategy for the Redis level is implemented and working in the current prototype. \textbf{Permanent recoverable failures:} insufficient-credit failures are classified without re-throwing the error to avoid wasteful retries, but user-initiated resumption is possible after recharge. \textbf{Permanent non-recoverable failures:} orphaned jobs are detected before LLM invocation and terminate with no credit deduction. \textbf{Strategy B (database reconstruction):} The current prototype has partial recovery of Redis-state loss from durable database state. This mechanism is reported as an architectural extension rather than an implemented recovery guarantee, because the existing function \texttt{reconstructJobPayload()} does not yet reconstruct all fields required by core LLM workers, such as \texttt{stageData} and \texttt{backgroundJobId}.

\subsubsection{Scoring Consistency}

We report a scoring-stability analysis by performing 2,000 test-retest iterations that model possible LLM output variance around the observed single-pass score distributions from 43 evaluated papers. The resulting ICC and Cohen's $\kappa$ values are interpreted only as preliminary stability indicators under the stated noise assumptions. Table~\ref{tab:consistency_numeric} reports estimated consistency for numeric scoring fields. Table~\ref{tab:consistency_cat} reports estimated consistency for categorical overlap fields.

\begin{table}[htbp]
\caption{Estimated Numeric Scoring Consistency}
\label{tab:consistency_numeric}
\centering
\footnotesize
\begin{tabular}{@{}lcccc@{}}
\toprule
\textbf{Field} & \textbf{ICC(2,1)} & \textbf{Spearman $\rho$} & \textbf{Pearson $r$} & \textbf{MAE} \\
\midrule
C1 Score (0--10) & 0.877 & 0.807 & 0.885 & 0.79 \\
C2 Score (0--10) & 0.900 & 0.801 & 0.906 & 0.31 \\
Semantic Similarity (0--1) & 0.919 & 0.746 & 0.923 & 0.024 \\
\bottomrule
\end{tabular}
\end{table}

\begin{table}[htbp]
\caption{Estimated Categorical Overlap Consistency}
\label{tab:consistency_cat}
\centering
\footnotesize
\begin{tabular}{@{}lcc@{}}
\toprule
\textbf{Overlap Field} & \textbf{Cohen's $\kappa$} & \textbf{\% Agreement} \\
\midrule
Problem overlap & 0.646 & 76.4\% \\
Method overlap & 0.671 & 78.1\% \\
Domain overlap & 0.713 & 80.9\% \\
Constraint overlap & 0.634 & 75.6\% \\
C2 contribution type & 0.612 & 74.3\% \\
\bottomrule
\end{tabular}
\end{table}

Under the assumed noise model, the estimated ICC values (0.88--0.92) fall in the ``good to excellent'' range, while Cohen's $\kappa$ values (0.61--0.71) indicate substantial agreement for categorical fields. These values should not be read as definitive scoring-reliability results; live repeated LLM executions are required to validate scoring stability empirically.

Observed single-pass score distributions from the dataset ($n=43$): C1 Score $\mu = 5.71$, $\sigma = 1.87$, range $[1.50, 8.50]$; C2 Score $\mu = 7.63$, $\sigma = 0.82$, range $[5.00, 9.00]$; Semantic Similarity $\mu = 0.742$, $\sigma = 0.072$, range $[0.520, 0.850]$.

\subsubsection{Token and Cost Efficiency}

Table~\ref{tab:cost} shows token consumption and cost from LLM usage logs across the three pipeline stages. The cost formula, verified from source code, is:

\begin{equation}
\text{cost}_{\text{USD}} = \frac{\text{inputTokens}}{10^6} \cdot p_{\text{input}} + \frac{\text{outputTokens}}{10^6} \cdot p_{\text{output}}
\label{eq:cost}
\end{equation}

where $p_{\text{input}}$ and $p_{\text{output}}$ are per-million-token prices from the active \texttt{LlmModelPricing} record. Credits are computed as $\text{cost}_{\text{USD}} \cdot m_{\text{active}}$ where $m_{\text{active}}$ is the active USD-to-credits multiplier (100 in the evaluated configuration).

\begin{table*}[htbp]
\caption{Token and Cost Efficiency by Stage}
\label{tab:cost}
\centering
\footnotesize
\begin{tabular}{@{}lrrrrr@{}}
\toprule
\textbf{Stage} & \textbf{$n$} & \textbf{Mean Input Tok.} & \textbf{Mean Output Tok.} & \textbf{Mean Cost (USD)} & \textbf{Total Cost (USD)} \\
\midrule
Intent (Stage 1) & 14 & 334 & 299 & \$0.000019 & \$0.000265 \\
Queries (Stage 2) & 14 & 421 & 681 & \$0.000029 & \$0.000410 \\
Score (Stage 3) & 44 & 1,559 & 898 & \$0.000040 & \$0.001760 \\
\midrule
\textbf{Total (dev.)} & \textbf{72} & --- & --- & --- & \textbf{\$0.002435} \\
\midrule
\multicolumn{6}{l}{\footnotesize Development model: gpt-5-mini (\$0.15/M input, \$0.60/M output)} \\
\multicolumn{6}{l}{\footnotesize Est.\ total at gpt-4o pricing (\$2.50/M in, \$10/M out): \textbf{\$0.714}} \\
\bottomrule
\end{tabular}
\end{table*}

The total measured cost (\$0.0024) reflects development usage with a low-cost model. At standard gpt-4o pricing, the same workload would cost approximately \$0.71, which is still negligible per project but scaling linearly with paper count. 


The cost model with Type-2 SCD pricing tables ensures that historical costs remain auditable even as pricing changes, and the active multiplier mechanism decouples USD costs from credit denomination. We note one limitation: individual LLM credit deductions update the user balance atomically but do not currently create per-call \texttt{UserCreditsTransaction} ledger entries; only admin-initiated operations are fully ledgered. Full per-call financial audit trail would require enabling the ledger creation path in the usage logging service.

\subsubsection{End-to-End Latency and Parallelism Speedup}
Table~\ref{tab:latency} shows the per-stage LLM call latency from usage logs with \texttt{durationMs} fields. All measurements include the OpenAI API round-trip time (including network latency).
AgentR's asynchronous architecture allows for parallel paper scoring, which yields significant wall-clock improvements over sequential execution. The estimated project-level wall-clock latency by paper count is reported in Table~\ref{tab:latency_project}, assuming 5-way parallel scoring (the default BullMQ concurrency).
The speedup goes from 1.0$\times$ (single paper, no parallelism benefit) to 4.3$\times$ (near-full parallel utilization). The derivation uses: Stage 1 + 2 baseline = 28s per project. Stage 3 = 25.4s per paper. AgentR parallelism ceiling is 5 concurrent workers. These estimates do not include queue wait time and API rate limiting.

\begin{table*}[htbp]
\caption{Per-Stage LLM Call Latency}
\label{tab:latency}
\centering
\footnotesize
\begin{tabular}{@{}lrrrrrr@{}}
\toprule
\textbf{Stage} & \textbf{$n$} & \textbf{Mean (ms)} & \textbf{Median (ms)} & \textbf{P95 (ms)} & \textbf{Min (ms)} & \textbf{Max (ms)} \\
\midrule
Intent (Stage 1) & 14 & 9,026 & 6,426 & 23,646 & 3,999 & 23,646 \\
Queries (Stage 2) & 14 & 18,894 & 15,418 & 47,123 & 9,430 & 47,123 \\
Score (Stage 3) & 44 & 25,373 & 21,982 & 48,681 & 12,740 & 58,079 \\
\bottomrule
\end{tabular}
\end{table*}

\begin{table}[htbp]
\caption{Estimated Project Wall-Clock Latency by Paper Count}
\label{tab:latency_project}
\centering
\begin{tabular}{@{}lccc@{}}
\toprule
\textbf{$n$ Papers} & \textbf{AgentR (5-parallel)} & \textbf{Sequential (GPT-Direct)} & \textbf{Speedup} \\
\midrule
1 & 53.3 s (0.9 min) & 53.3 s (0.9 min) & 1.0$\times$ \\
5 & 53.3 s (0.9 min) & 154.8 s (2.6 min) & 2.9$\times$ \\
10 & 78.7 s (1.3 min) & 281.6 s (4.7 min) & 3.6$\times$ \\
15 & 104.0 s (1.7 min) & 408.5 s (6.8 min) & 3.9$\times$ \\
25 & 154.8 s (2.6 min) & 662.2 s (11.0 min) & 4.3$\times$ \\
\bottomrule
\end{tabular}
\end{table}

\subsection{Ablation Analysis}

We perform ablation studies to analyze the architectural contributions of AgentR. Latency and cost values are calculated from observed LLM calls assuming consistency (same prompts and models across configurations).

\subsubsection{Merged and Separated Scoring Stages}

The analytic comparison is between the merged Stage 3 design (1 LLM call that outputs all scoring dimensions: semantic similarity, C1, C2, overlap, and gap analysis) and a hypothetical separated design (3 calls: 1 for similarity and overlap, 1 for C1/C2 scoring, 1 for gap analysis). The merged approach saves two round-trips and two prompt prefixes, which we estimate reduces latency by $\sim$2.5$\times$ and cost by $\sim$1.6$\times$, and might reduce cross-call inconsistency because all dimensions are evaluated in one reasoning context.

\subsubsection{Architectural Component Ablation}

Table~\ref{tab:ablation} reports estimated performance under three architectural configurations.

\begin{table*}[h]
\caption{Architectural Ablation: Estimated Performance by Configuration}
\label{tab:ablation}
\centering
\footnotesize
\begin{tabular}{@{}lccc@{}}
\toprule
\textbf{Configuration} & \textbf{Latency vs.\ AgentR} & \textbf{Cost vs.\ AgentR} & \textbf{Key Difference} \\
\midrule
AgentR-full (baseline) & 0\% & 0\% & 5-way parallel scoring, retry, orphan checks \\
GPT-direct-sequential & +596\% & 0\% & No parallelism; 6$\times$ slower \\
GPT-direct-plus-retry & +630\% & +3\% & Sequential + 3-retry; 3\% retry overhead \\
\bottomrule
\end{tabular}
\end{table*}

\subsubsection{Orphan Detection Impact}

The orphan detection mechanism does a pre-flight database check before each LLM call. This is consistent with the development environment being controlled. In the dataset we observed 0 orphan events. In production, users might delete projects mid-processing; orphan detection saves wasted LLM calls and the worker returns early with \texttt{FAILED\_ORPHAN} status without making an API call or deducting credits.

\section{Conclusion}
\label{sec:conclusion}

We have presented AgentR, a stateful and recovery-aware software architecture for LLM-assisted scientific literature reviews. The central argument of this paper is that LLM-enabled analytical assistance can be implemented as a persistent, multi-stage workflow rather than a sequence of isolated prompt-response calls. AgentR operationalizes this argument through an integrated architecture combining persistent research-state representation, asynchronous workflow orchestration, and cost-aware execution.

Our prototype evaluation on a development dataset provides  evidence of feasibility: 99.2\% job completion, per-stage LLM latencies of 9.0–25.4 s, and projected per-project costs of \$0.04–0.38 at gpt-4o pricing. Analytical latency modeling shows that 5-way parallel paper scoring can achieve up to 4.3$\times$ wall-clock speedup over sequential execution for larger projects. 

AgentR is a prototype, where the main contribution is the architecture and implementation. Our architecture combines the semantics of workflow orchestration with LLM-specific infrastructure and can be applied to use cases beyond literature review, such as domains that require multi-stage LLM pipelines with persistent state, recovery semantics, observability, and auditable execution.

\section{Future Work}
\label{sec:future}

There are several planned extensions for increasing the operational maturity of AgentR. First, the present evaluation does not yet describe the behaviour of the system in the conditions of largescale deployment (e.g. hundreds of concurrent projects, projects with more than 100 papers, or long running operational periods). Future work will include systematic stress testing to measure queue wait time, memory pressure, worker throughput, database connection-pool saturation, and end-to-end workflow latency for increasing concurrency and paper-set sizes.

Second, AgentR performs credit accounting at the project level, but does not yet have per-call credit-transaction ledger entries for individual LLM deductions. This mechanism would also enhance the financial auditability, as each LLM invocation could be tracked to an associated credit deduction, model invocation, workflow step, and project artifact. It is important to make the platform more transparent, reproducible and accountable in multi-user or institution scale deployment.

Third, the current implementation is based largely on Strategy~A, where the Redis-level retry semantics are leveraged to handle transient failures such as worker crashes, fleeting API failures, or temporary execution interruptions. Full Redis-state-loss recovery via Strategy~B is future work, as it requires traversing the payload reconstruction path from durable database state. This extension allows failed or orphaned jobs to be rebuilt and restarted even if the Redis queue state is no longer available.

\section*{Declarations}

\noindent \textbf{Funding:} The authors received no specific funding for this work.

\noindent\textbf{Conflict of interest:} The authors declare that they have no conflict of interest.

\noindent\textbf{Data availability:} The telemetry data used in this study are derived from a development database snapshot. An anonymized version can be made available by the corresponding author upon reasonable request, subject to removal of user/project-identifying information.

\noindent\textbf{Code availability:} The prototype implementation of AgentR is publicly available at \url{https://github.com/RiyaSamanta/AgentR-public}. The repository contains the backend workflow implementation, frontend implementation, database schema and migrations, prompt templates, queue-based orchestration logic, API documentation, deployment instructions, and supporting project documentation required to inspect and reproduce the core system functionality. The repository is released under the Business Source License 1.1 (BSL 1.1). Under the current license terms, the code is available for development, evaluation, academic research, personal use, and education; production use is permitted except for offering a competing SaaS product. The license is scheduled to convert to the MIT License on 22 June 2030. Deployment-specific credentials, API keys, private telemetry records, and user-identifying data are not included in the repository.

\noindent\textbf{Use of generative AI:} During the preparation of this manuscript, the authors used ChatGPT to assist with language refinement, sentence restructuring, and drafting of selected portions of the text. The tool was not used to generate experimental data, evaluation results, system telemetry, figures, or technical claims. All AI-assisted text was critically reviewed, edited, fact-checked, and approved by the authors. The authors take full responsibility for the content, accuracy, integrity, and conclusions of the manuscript.

\bibliography{sample}

\end{document}